\documentclass[twocolumn,showpacs,preprintnumbers,amsmath,amssymb]{revtex4}
\usepackage{graphics}
\usepackage{epsfig}
\usepackage{color}
\usepackage{subfigure}  

\begin{document}

\title{Angular Distribution of Nonlinear Harmonic Generation in Helical Undulators: a comparison between experiments and theory}
\author{E. Allaria$^{1}$, G. De Ninno$^{1,2}$, G. Geloni$^{3}$, C. Spezzani$^{1}$}
\affiliation{1. Sincrotrone Trieste, S.S. 14 km 163.5, Basovizza (Ts), Italy \\
2. Physics Department, Nova Gorica University, Nova Gorica, Slovenia\\
3. European XFEL GmbH, Albert-Einstein-Ring 19, 22761 Hamburg, Germany\\
}

\date{\today}

\begin{abstract}
Using harmonic emission from circularly polarized undulator is a procedure 
that is normally employed on synchrotron beamlines in order to extend the 
covered spectral range. A similar capability is likewise beneficial 
for next generation free-electron lasers. In this paper, we perform a first 
quantitative experimental analysis of the angular distribution of free-electron 
laser harmonic emission from helical undulators. Experimental results are compared 
to those obtained by means of a theoretical model based on the paraxial solution 
of Maxwell's equations. 
\end{abstract} 

\maketitle

\section{Introduction}
Variable polarization is one of the most attractive features of light 
sources based on electron beams propagating through APPLE-type undulators. 

Such a feature is particularly attractive for those applications aiming to investigate the local symmetry of a given system, especially when the symmetry can be ascribed to the lattice geometry (as in the case of crystals), to the chirality of a molecule or to a possible local atomic magnetic moment. In fact, several spectroscopic methods rely on the opportunity to chose a well defined polarization state of the light. Spin \cite{pescia} and angular resolved photo-electron spectroscopy, x-ray linear and circular dichroism or resonant scattering of polarized x-rays \cite{blume, altarelli} are only few examples.
For this reason, light polarization-dependent spectroscopy using standard synchrotron radiation 
has become a very powerful tool to study the electronic and magnetic 
properties of matter \cite{sut, blume2}. 

Thanks to a dramatic increase of photon peak-brilliance with respect to 
conventional sources, as well 
as to the possibility of controlling both the temporal duration and the 
spectral bandwidth of the produced light pulses, free-electron lasers 
(FELs) will allow completely new studies in the above mentioned research fields \cite{appl1}. Such studies will also take advantage 
of well-defined and easily tunable light polarization. 
For the preparation of the scientific case of single-pass FELs, a 
fundamental question concerns the possibility of generating significant 
photon flux at higher harmonics of the fundamental wavelength, while 
maintaining polarization ductility. 

In a FEL, harmonic generation is driven by the electron-beam interaction 
with a radiation at a given fundamental wavelength, in the presence of the static and periodic 
magnetic field generated by an 
undulator \cite{Schmitt86,Bonifacio90,pino1,Huang00,Biedron02,Freund05,Saldin06, pino2}. 
Such interaction produces a significant modulation (called bunching)
of the electron-beam density at the fundamental wavelength, and  
its harmonics. Harmonic bunching is the (nonlinear) source of FEL harmonic 
emission. The latter is orders of magnitudes stronger than the spontaneous 
incoherent emission and provides the possibility to extend the FEL tuning 
range towards short wavelengths. 

In the simplest possible configuration, FEL harmonics are produced in 
the same undulator where radiation at the fundamental wavelength 
is generated. This method is the standard one when the FEL light is 
obtained from Self Amplified Spontaneous Emission (SASE) \cite{sase1}, 
and is normally referred to as nonlinear harmonic generation (NHG). 

An alternative approach to NHG for producing coherent harmonic emission 
relies upon the bunching created in an undulator, called the modulator, 
for generating coherent emission in a subsequent undulator, called the 
radiator, tuned at one of the harmonics of the modulator \cite{yu1, yu2}. 
This technique is usually called coherent harmonic generation 
(CHG). The implementation of CHG requires the use of an external coherent 
source, acting as a ``seed'', which is normally provided by one of the
harmonics of a Ti: sapphire laser. These harmonics can be produced either 
using a solid-state crystal, or a gas jet \cite{hhg}. The seed-electron 
interaction in the modulator is necessary to initiate the bunching process. 
Bunching is produced at all (odd and even) harmonics. As a consequence, 
in the radiator, coherent emission may occur at any selected harmonics, 
no matter the radiator polarization. 
The situation changes when NHG is considered. It is well known that, 
in case of planar undulators, on-axis NHG occurs only at odd harmonics, 
while even harmonics are present only off-axis \cite{Schmitt86}. 
In \cite{prl1} it has been experimentally demonstrated that, as in the case of spontaneous emission, the NHG signal 
generated by helical undulators is distributed off-axis.
A quantitative estimate of the off-axis coherent harmonic flux is, however, still missing. 
This lack of information is due to the absence, so far, of an experimental 
characterization of the angular distribution of harmonic emission and, at 
the same time, to the difficulty of modelling the coherent emission from 
a modulated electron beam in the far zone. Recently, an analytical model 
has been proposed \cite{NHGP,OURF,OURI,NHGH}, relying on the paraxial 
solution of Maxwell's equations. Such a model confirms what has been 
found in \cite{prl1}, i.e. no on-axis NHG in helical undulators, and 
provide a quantitative prediction of the far-field angular harmonic distribution. 

In this paper, we perform a first quantitative experimental analysis of 
the angular distribution of FEL harmonic radiation generated by a helical 
undulator. In particular, we characterize the off-axis emission in NHG 
configuration. Obtained results are used to benchmark the theoretical 
model proposed in \cite{NHGP,OURF,OURI,NHGH}. 

The paper is structured as follows. In Section \ref{sect2} we briefly 
review the theoretical model and discuss the angular characteristics 
of the predicted far-field harmonic distribution. In Section \ref{sect3} 
we present the experimental device we used to perform the measurements,
i.e. the FEL installed at the Elettra storage-ring \cite{prl2, nim}. 
In Section \ref{sect4}, experimental data are presented and compared 
to theoretical predictions. Finally, in Section \ref{sect5} we draw 
conclusions and provide perspectives for future work. 
       
\section{Theoretical framework}
\label{sect2}
The coherent emission in the far zone from modulated electron
beams in planar or helical undulators can be studied by solving the
paraxial Maxwell's equation for the electric field in the
space-frequency domain \cite{NHGP,OURF,OURI,NHGH}.

\begin{figure}[t]
\resizebox{0.48\textwidth}{!}{\includegraphics{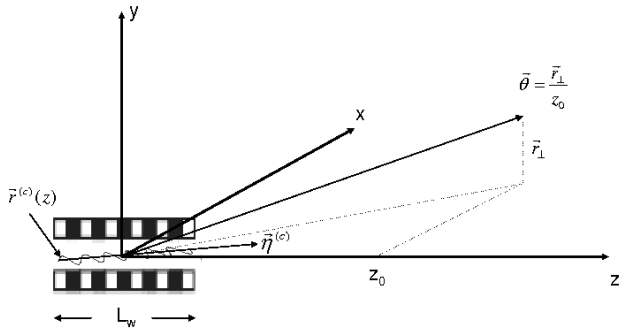}}
\caption{Reference system and symbols used to develop the theoretical model. \label{fig1}}
\end{figure}

Let us call the transverse electric field in the space-frequency
domain $\vec{\bar{E}}_\bot (z,\vec{r}_\bot,\omega)$, where, as shown in Fig. \ref{fig1},
$\vec{r}_\bot = x \vec{e}_x+y\vec{e}_y$ identifies a point on a
transverse plane at longitudinal position $z$, $\vec{e}_x$ and
$\vec{e}_y$ being unit vectors in the transverse $x$ and $y$
directions. 

The paraxial approximation is well suited for modelling the emission from ultra-relativistic electrons (see
\cite{OURF,OURI}). It relies on the assumption that the electric field envelope
$\vec{\widetilde{E}}_\bot = \vec{\bar{E}}_\bot \exp{[-i\omega z/c]}$
does not vary much along $z$ on the scale of the reduced
wavelength $\lambdabar=\lambda/(2\pi)$. The wave equation for the
slowly-varying envelope of the electric field is found to be 

\begin{eqnarray} \left({\nabla_\bot}^2 + {2 i \omega \over{c}}
{\partial\over{\partial z}}\right)
\left[\vec{\widetilde{E}}_\bot(z,\vec{r}_\bot,\omega)\right] = \nonumber\\ 
- {4 \pi} \left(\frac{i\omega}{c^2}\vec{\bar{j}}_\bot
-\vec{\nabla}_\bot \bar{\rho}\right) \exp\left[-\frac{i \omega
z}{c}\right] ~ .\label{field1}
\end{eqnarray}
(where $c$ is the speed of light in vacuum). The right hand side is determined by the trajectory of the source
electrons, and is written in terms of the Fourier transform of the
transverse current density, $\vec{\bar{j}}_\bot
(z,\vec{r}_\bot,\omega)$, and of the charge density,
$\bar{\rho}(z,\vec{r}_\bot,\omega)$, which will be considered as
given macroscopic quantities. The Fourier transform of the charge density can
be written as

\begin{eqnarray}
\bar{\rho} = - \widetilde{\rho}(z,\vec{r}_\bot-\vec{r'}_{o
\bot}(z),\omega) \exp\left[
i\omega\frac{s_o(z)}{v_o}\right]~,\label{rhotr}
\end{eqnarray}
where the minus sign on the right hand side is introduced for
notational convenience. In the latter expression, the quantities $\vec{r'}_{o \bot}(z)$,
$s_o(z)$ and $v_o$ are, respectively, the transverse position, the curvilinear
abscissa and speed of a reference electron with Lorentz
factor $\gamma$ that is injected on axis with no deflection and is
guided by the undulator field. Such electron follows a trajectory
$\vec{r'}_{o\bot}(z)= r'_{ox} \vec{e}_x+r'_{oy} \vec{e}_y$, which
assumes different forms, namely:

\begin{eqnarray}
&&\vec{r'}_{o\bot}(z) = \frac{ K }{\gamma k_w} \left[(\cos(k_w z)-1)
\vec{e}_x + \sin(k_w z) \vec{e}_y\right]~ \label{rhelO}
\end{eqnarray}
if the electron propagates through a helical undulator and

\begin{eqnarray}
&&\vec{r'}_{o\bot}(z) = \frac{ K }{\gamma k_w} (\cos(k_w z)-1)
\vec{e}_x ~ \label{rhelp}
\end{eqnarray}
if the undulator is planar. Here $K=\lambda_w e H_w/(2\pi m_e c^2)$ is
the undulator parameter, $\lambda_w = 2\pi/k_w$ being the
undulator period, $e$ the electron charge, $H_w$ the
maximal modulus of the on-axis magnetic field of the undulator, and $m_e$
the rest mass of the electron. The corresponding velocity is
given by $\vec{v}_{o\bot}(z)= v_{ox} \vec{e}_x+v_{oy}
\vec{e}_y$.

Since we want to discuss an FEL process, $\widetilde{\rho}$ in Eq.
(\ref{rhotr}) is a slowly varying function of $z$ on the
wavelength scale and is peaked around the harmonics of the
fundamental (i.e. at frequencies $ h \omega_r = 2 h k_w c
\bar{\gamma}_z^2$, $\bar{\gamma}_z$ being the average longitudinal
Lorentz factor, and $h$ the harmonic number) with bandwidth
$\Delta \omega/(h \omega_r) \ll 1$.

Moreover, for each particle in the beam, the relative deviation of the particles energy from $\gamma m_e c^2$ is small, i.e. 
$(\gamma-\bar{\gamma}) /\gamma \ll 1$.
Therefore, in first approximation, we can neglect the difference between the average
transverse velocity of electrons $\langle \vec{v}_\bot \rangle$
and $\vec{v}_{o\bot}$, so that $\vec{\bar{j}}_\bot \equiv \langle
\vec{v}_\bot \rangle \bar{\rho} \simeq \vec{v}_{o\bot}
\bar{\rho}$.

Furthermore, we will be interested in the total power emitted and
in the directivity diagram of the radiation in the far zone. We
therefore introduce the far zone approximation calling the
observation angle $\vec{\theta}=\vec{r}_{\bot o}/z_o$ (see Fig.\ref{fig1}), $z_0$ being
the distance from the middle of the undulator to the observer. We also set
$\theta \equiv |\vec{\theta}|$ and take the limit for
$z_o \gg L_w$, where $L_w = N_w \lambda_w$ is the undulator
length.

Finally, we introduce a coherent deflection angle,
$\vec{\eta}^{(c)}$, to describe the transverse deflection of the electron
beam as a whole \footnote{With this, we assume that the deflection
angle $\vec{\eta}^{(c)}$ is constant. This is the case only if we
do not account for focusing elements within the undulator.}. We
therefore perform the following substitution:

\begin{eqnarray}
&&\vec{r'}_{o}(z) \longrightarrow
\vec{r}^{(c)}(z,\vec{\eta}^{(c)}) = \vec{r'}_{o}(z) +
\vec{\eta}^{(c)} z ~. \label{etac}
\end{eqnarray}
It follows that also $\vec{v}_{o}(z)$ should be substituted with
$\vec{v}(z,\vec{\eta}^{(c)})$, and that
$\bar{\gamma}_z(z,\vec{\eta}^{(c)})$ is now a function of both $z$ and
$\vec{\eta}^{(c)}$. With these prescriptions, we obtain the
following solution of Eq. (\ref{field1}):

\begin{eqnarray}
&&\vec{\widetilde{E}}_{\bot }= \nonumber\\ 
&-&\frac{i \omega }{c z_o}
\int_{-\infty}^{\infty} dz' \int d \vec{l}'~
\left(\frac{\vec{v}_\bot(z',\vec{\eta^{(c)}})}{c}
-\vec{\theta}\right) \widetilde{\rho} \left(z',\vec{l}',\omega
\right)\nonumber\\
&&\times \exp\left\{\frac{i\omega}{2c} \left[z_o
\theta^2 - 2 \vec{\theta}\cdot
\vec{r}^{(c)}_\bot(z',\vec{\eta^{(c)}})- 2 \vec{\theta}\cdot
\vec{l}' +z'\theta^2\right]\right\} \nonumber\\
&&\times\exp\left\{+ i \int_{0}^{z'} d \bar{z}\frac{
\omega }{2
 c\bar{\gamma}_z^2(\bar{z},\vec{\eta}^{(c)})}\right\}~
.\cr && \label{generalfin3}
\end{eqnarray}
Substitution of the equations of motion Eq. (\ref{rhelO}) or Eq.
(\ref{rhelp}), and of an appropriate expression for $\widetilde{\rho}$ into Eq.
(\ref{generalfin3}), yields the electric field in the far zone. We
will consider a simple model where

\begin{eqnarray}
&&\widetilde{\rho}\left(z',\vec{l}',\omega \right) = \frac{(-e) N_e
\bar{f}(\omega)}{2\pi \sigma_x\sigma_y}\nonumber\\
&&
\exp\left[-\frac{l_x^{'2}}{2\sigma_x^2}\right]
\exp\left[-\frac{l_y^{'2}}{2\sigma_y^2}\right]\exp\left[i
\frac{\omega}{c} \vec{\eta}^{(c)}\cdot
\vec{l}'\right]~,\label{rhotilde}
\end{eqnarray}
$\sigma_{x,y}$ being the electron beam rms transverse size, and
$\bar{f}(\omega)$ the Fourier transform of the bunch
temporal profile:

\begin{eqnarray}
f(t) = \frac{I(t)}{(-e)N_e}\left[1+b_h\cos(h\omega_r t)\right]~.
\label{ft}\end{eqnarray}
Here $b_h$ is the bunching at the $h$-th harmonic \footnote{Note that this definition of bunching gives a coefficient 
which is a factor two larger than the one used in other works (see, e.g., \cite{yu2}).}.

This model describes a bunched electron beam with
Gaussian transverse profile, with current $I(t)$,
longitudinally modulated at frequency $h \omega_r$. We are 
interested in a frequency range near the fundamental harmonic,
$\omega_r$, or its $h$-th integer multiple. We specify how close
the frequency $\omega$ is to the $h$-th harmonic by defining a
detuning parameter $C_h$:

\begin{eqnarray}
C_h = \frac{\omega}{2\bar{\gamma}_z^2 c} - h k_w = \frac{\Delta
\omega}{\omega_r} k_w~, \label{Ch}
\end{eqnarray}
where $\omega = h \omega_r + \Delta \omega$. Whenever

\begin{equation}
C_h +\frac{\omega}{2 c}\left[
\left(\theta_x-\eta_x^{(c)}\right)^2+\left(\theta_y-\eta_y^{(c)}\right)^2\right]
\ll k_w \label{eqq} ~,
\end{equation}
and $N_w \gg 1$, major simplifications arise because fast
oscillating terms in powers of $\exp[i k_w z']$, which explicitly
appear in Eq. (\ref{generalfin3}) after substituting the
trajectory \cite{ALFE}, effectively
average to zero after integration in $dz'$. To complete the
resonance approximation, we further select frequencies such that

\begin{eqnarray}
\frac{|\Delta \omega|}{\omega_r} \ll 1~.
\label{resext}
\end{eqnarray}
This condition on frequencies automatically selects the observation
angles of interest: $h (\vec{\theta}-\vec{\eta}^{(c)})^2 \ll
1/\bar{\gamma}_z^2$. Under the constraint imposed by
(\ref{resext}), independently of the value of $K$, we obtain from Eq.
(\ref{generalfin3}) analytical expressions for the angular spectral
fluxes\footnote{In CGS units. Conversion to SI units is done by dividing by a factor of $10^{-7}$.}. These expressions describe the coherent
emission in the far zone from modulated electron beams in planar
or helical undulators, and strongly depend on the Fresnel
parameter $N_{x,y}$ defined as

\begin{eqnarray}
N_{x,y} = \frac{h \omega_r\sigma_{x,y}^2}{c L_w} \label{Nxy}~.
\end{eqnarray}
In the helical case, the spectral flux at the $h$-th
harmonic is given by:

\begin{eqnarray}
&&\frac{dP_h}{d\Omega} = \frac{\pi}{c} \left[\frac{b_h I(t) h
 N_w K \bar{\gamma}_z}{((h-1)!)(1+K^2)}\right]^2\nonumber\\
&& \left[\frac{h^2 L_w K^2 \omega_r}{4 \pi N_w
c (1+K^2)}\left(\theta_x^2+\theta_y^2\right)\right]^{h-1} \nonumber\\
&&\times \mathrm{sinc}^2\left[\frac{h L_w \omega_r}{4 c}
\left(\theta_x^2+\theta_y^2\right)\right] \nonumber\\ 
&&\times \exp\left[-\frac{h L_w
\omega_r}{c} \left(N_x \theta_x^2+N_y \theta_y^2\right)\right]~.
\label{helical}
\end{eqnarray}

\begin{figure}[t]
\resizebox{0.48\textwidth}{!}{\includegraphics{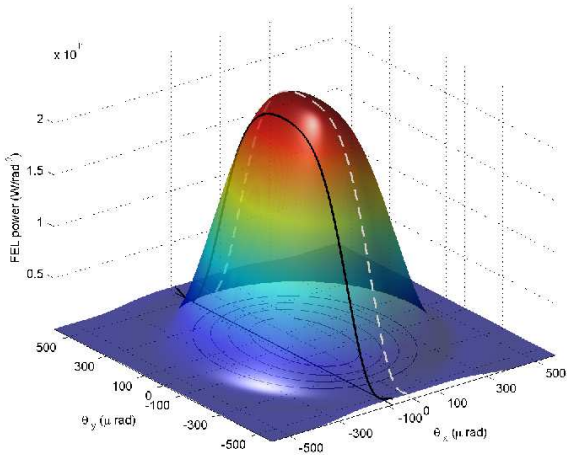}}
\caption{Angular spectral flux in CHG mode, as predicted by Eq. (\ref{helical}). 
For the calculation, use has been made of the parameters reported in Table \ref{table1}, 
corresponding to the case of the Elettra storage-ring FEL.
In this case, light is emitted at the radiator fundamental wavelength (i.e., $h=1$, corresponding to 260 nm), which is the third harmonic of the seed wavelength (780 nm).
\label{fig2}.
}
\end{figure}

\begin{figure}[t]
\resizebox{0.48\textwidth}{!}{\includegraphics{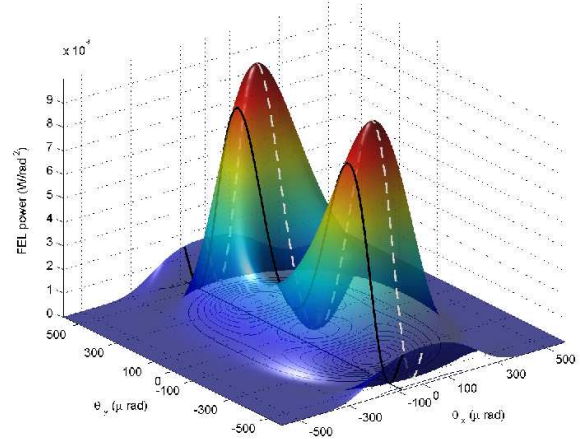}}
\caption{Angular spectral flux in NHG mode, as predicted by Eq. (\ref{helical}). 
For the calculation, use has been made of parameters reported in Table \ref{table1}, 
corresponding to the case of the Elettra storage-ring FEL. 
In this case, light is emitted at the second harmonic of the radiator fundamental wavelength (i.e., $h=2$, corresponding to 260 nm), which is the third harmonic of the seed wavelength (780 nm).
\label{fig3}}
\end{figure}

Figures \ref{fig2} and \ref{fig3} show the angular spectral flux as 
obtained from Eq. (\ref{helical}) for the CHG and NHG, respectively. 
For the calculation, use has been made of the parameters reported in 
Table \ref{table1}, corresponding to the experimental setup described 
in the following Section. 

\section{Experimental setup}
\label{sect3}

\begin{figure}[t]
\resizebox{0.48\textwidth}{!}{\includegraphics{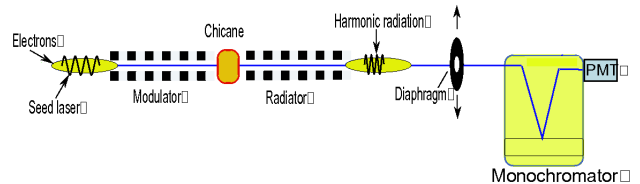}}
\caption{Schematic layout of the Elettra storage-ring FEL.\label{fig4}}
\end{figure}

The predictions of the model presented in the previous Section have been tested on 
the Elettra storage-ring FEL, whose scheme is shown in Fig. \ref{fig4}. The setup, 
which is based on two independent APPLE II undulators separated by a magnetic 
chicane, is very flexible: depending on the undulator configuration, both NHG 
and CHG schemes can be implemented. The laser pulse emitted by a Ti:sapphire 
laser (fundamental wavelength at 780 nm) interacts with the electron bunch 
within the modulator, inducing an energy modulation in the electron beam. After 
the conversion of the energy modulation into spatial bunching, occurring 
into the magnetic chicane, the bunch enters the radiator and emits coherent 
radiation at one of the harmonics of the seed laser (the third one, i.e. 260 nm, 
for the experiments presented in this paper). The harmonic radiation is first 
monochromatized and then acquired on a digital oscilloscope by means of a 
photomultiplier tube (PMT), see Fig. \ref{fig4}. The latter is fast enough to 
resolve the pulsed dynamics of the electron bunch photoemission (nanonosecond scale),
but does not allow to resolve the photon pulse shape (tens of ps). 
A diaphragm of 1.4 mm is placed downstream the radiator, before the detector. 
The diaphragm is used to select only the portion of radiation emitted within 
a small cone close to the undulator axis (angular acceptance of about 100 $\mu$rad). 
By moving the diaphragm, it is possible to characterize the angular distribution 
of the produced coherent radiation. 

For the reported experiments, Elettra was operated in single-bunch mode 
at 0.75 GeV ($\gamma \simeq 1468$), the peak bunch current and 
its revolution period being, respectively, $\simeq 10$ A and 864 ns. 
The seed repetition rate is 300 Hz. The laser pulse and the electron 
bunch duration are, respectively, 120 fs (rms) and about 35 ps (rms). 
A stable laser-electron synchronization with a jitter of few ps \cite{nim} 
guarantees an efficient and stable seeding process, with shot-to-shot fluctuations 
of the order of few percent. The main experimental parameters, also used to obtain 
the theoretical angular spectral flux reported in Figs. \ref{fig2} and \ref{fig3}, 
are reported in Table \ref{table1}. More detailed information about the 
experimental setup, as well as about the performance of the Elettra 
storage-ring FEL, can be found in \cite{nim}. 

\begin{table}
\caption{\label{table1} Main experimental parameters.}
\begin{ruledtabular}
\begin{tabular}{lll}
\bf{Electron beam}& & \\
\hline
Peak current & 10 & A \\
Normaized energy ($\gamma$) & 1468 &  \\
Bunch length (rms) & 35 & ps \\
Relative energy spread	& $\simeq 0.1$	 & \%	\\
Bunching ($b_h$)&0.6& \\
Horizontal transverse size (rms) ($\sigma_x$) & 200 & $\mu$m	\\ 
Vertical transverse size (rms) ($\sigma_y$) & 10 & $\mu$m\\ 
\hline
\bf{Seed laser}& & \\
\hline
Power & $\sim 10$ & GW	\\
Wavelength & 780 & nm	\\
Pulse length (rms) & $\sim$ 120	 & fs\\
Polarization	& Horiz., Circ. & \\
\hline
\bf{Radiator}& & \\
\hline
Periodicity & 0.10 & m \\
Number of periods ($N_w$) & 20 & \\
Resonant wavelength ($\displaystyle{\frac{2\pi c}{\omega_r}}$), CHG mode & 260 & nm \\
Undulator parameter ($K$) in CHG mode & 3.19 & \\
Resonant wavelength ($\displaystyle{\frac{2\pi c}{\omega_r}}$), NHG mode & 520 & nm \\
Undulator parameter ($K$) in NHG mode & 4.63 & \\ 
Polarization & Horiz, Circ. & \\
\end{tabular}
\end{ruledtabular}
\end{table}

\section{Results}
\label{sect4}

In this Section we present the experimental data and perform a quantitative 
comparison between experiments and theory. 

\begin{figure}[t]
\resizebox{0.48\textwidth}{!}{\includegraphics{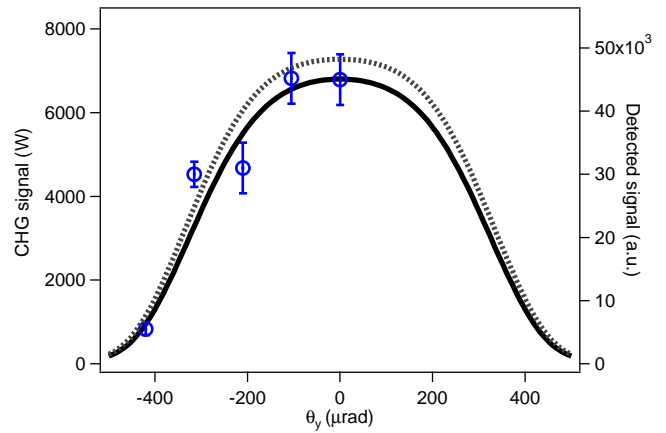}}
\caption{Intensity of the CHG signal at 260 nm, third harmonic of the
seed wavelength, as a function of the angle $\theta_y$ (vertical direction, 
see Fig. \ref{fig2}) between the undulator axis and the line from the 
center of the ondulator to the position of the diaphragm. Open circles: 
measurements (right vertical scale). Continuous and dotted lines: 
theoretical predictions (left vertical scale) obtained using Eq. (\ref{helical}) 
and the set of parameters reported in Table \ref{table1}. Dotted line: 
integration of the curve shown in Fig.\ref{fig2} along $\theta_y$, over
an angle of 100 $\mu$rad around $\theta_x=0$. In plotting experimental
data, a correction factor has been used to take into account (slight) 
current variations occurred in the time period between successive 
measurements.\label{fig5}}
\end{figure}

A first set of measurements has been carried out in CHG configuration, tuning 
the radiator at 260 nm (third harmonic of the seed wavelength), in circular 
polarization. The intensity detected by the PMT, as a function 
of $\theta_y$, is represented by the open circles in Fig. \ref{fig5}. 
As expected (see Fig. \ref{fig2}), the intensity has a maximum in 
correspondence of the undulator axis ($\theta_y=0$). 

Due to some mechanical limitations intrinsic to our device, the
diaphragm can be moved only downstream along the vertical direction.
As a consequence, no measurements have been performed along 
positive $\theta_y$'s, nor along the $\theta_x$ direction. Error bars
represent the result of several independent measurements. Fluctuations
are mainly due to FEL instabilities and to possible undesired slight
shifts of the electron-beam trajectory in the time period between
successive measurements. Data have been acquired at quite low average
electron-beam current ($ < 1$ mA), that is in a regime characterized by 
relatively long beam lifetimes (order of hours). However, since the 
harmonic intensity varies as the square of the number of charges
involved into the emission process \cite{prl2}, even small current
variations may reflect in significant intensity fluctuations. In order
to take this effect into account, a correction factor has been applied
to experimental data. In Fig. \ref{fig5}, continuous and dotted lines
represent the theoretical predictions obtained using Eq. (\ref{helical})
and the parameters reported in Table \ref{table1}. The
dotted line has been obtained by integrating the curve shown in Fig.
\ref{fig2} along $\theta_y$, over an angle of 100 $\mu$rad 
(the diaphragm acceptance) around $\theta_x=0$ 
(corresponding to a perfect horizontal alignment between the electro
n beam and the undulator axis). 
The continuous line has been instead obtained by integrating the curve
shown in Fig. \ref{fig2} along $\theta_y$, over an angle of 100 $\mu$rad
around $\theta_x=80$ $\mu$rad. The latter value corresponds to the best
possible experimental accuracy in controlling the horizontal alignment
between the electron beam and the undulator axis.

The same (experimental and theoretical) analysis has been carried out
in NHG configuration. In order to reproduce as much as possible the
experimental conditions of the CHG configuration, the radiator has
been tuned at 520 nm (in circular polarization). This avoids any
seed-electron interaction inside the radiator and, as a consequence,
prevents any undesired degradation of the beam quality (i.e., an increase
of the beam energy spread).

\begin{figure}[t]
\resizebox{0.48\textwidth}{!}{\includegraphics{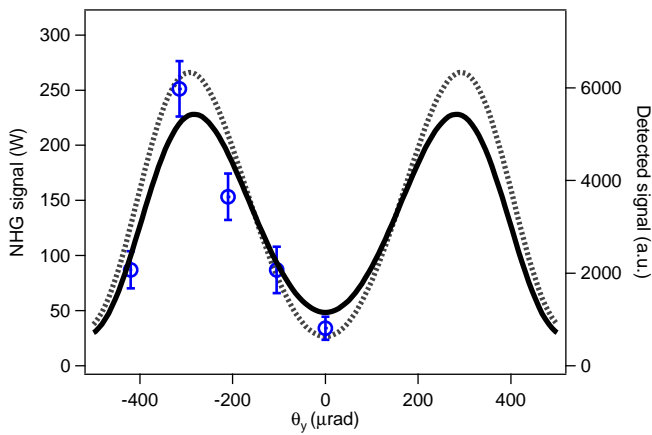}}
\caption{Intensity of NHG at 260 nm, second harmonic of the radiator
fundamental wavelength, as a function of the angle $\theta_y$. Open
circles: measurements (right vertical scale). Continuous and dotted
lines: theoretical predictions (left vertical scale) obtained using
Eq. (\ref{helical}) and the set of parameters reported in Table \ref{table1}.
Dotted line: cut of Fig. \ref{fig3} along $\theta_y$, for $\theta_x=0$;
continuous line: integration of the curve shown in Fig.\ref{fig2} along
$\theta_y$, over an angle of 100 $\mu$rad around $\theta_x=80$ $\mu$rad. 
In plotting experimental data, a correction factor has been used to take
into account (slight) current variations occurred in the time period
between successive measurements. \label{fig6}}
\end{figure}

Since 520 nm is not a harmonic of the input seed wavelength, 
no bunching is produced in the modulator at such wavelength. As a consequence, there is no coherent emission at the fundamental wavelength of the radiator.  However, one
obtains coherent emission through NHG at 260 nm and 130 nm (second and fourth harmonics, respectively, of the radiator wavelength). 
As for the case of CHG, the intensity of the coherent emission at 260 nm is detected
using the PMT. Results are reported in Fig. \ref{fig6}. As expected
(see Fig. \ref{fig3}), due to the finite aperture of the diaphragm,
NHG emission is not exactly zero on axis. 

In order to check the agreement between measurements and theory, one can
compare the experimental and theoretical ratios between the peak CHG and
residual NHG on-axis intensities. Using the data reported in Figs. \ref{fig5}
and \ref{fig6}, and considering the (more realistic) theoretical continuous
curves, one gets an experimental ratio of about 60 and a theoretical
one of 140. Taking into account the above-mentioned experimental incertitude,
as well as the approximations intrinsic to the theoretical model, the
agreement can be considered satisfactory \footnote{The agreement
becomes worse if the less realistic dashed theoretical curves in
Figs. \ref{fig5} and \ref{fig6} are used. In this case, one obtains a
theoretical ratio of about 280.}. 
 
\section{Conclusions}
\label{sect5}
We performed a quantitative analysis of the angular distribution of Nonlinear
Harmonic Generation in helical undulators. Measurements carried out on the
Elettra storage-ring free-electron laser, have been compared to the predictions
from a model based on the analytical solution of the paraxial Maxwell
equations. As we have shown, the agreement can be considered as satisfactory.
This result allows one to consider the theoretical model as a reliable tool
for predicting the off-axis photon flux of next-generation FELs, operated
in Nonlinear Harmonic Generation mode.

In fact, predictions are in general quite pessimistic. 
As an example, calculations have been recently performed~\cite{fawley} for harmonic off-axis emission in the case
of the FERMI@Elettra FEL \cite{fermicdr}, an FEL user facility presently under commissioning
at Sincrotrone Trieste.
In the case of FERMI, operating at 5 nm in circular polarization, FEL
saturation is easily reached and about $3\cdot 10^{12}$ 
on-axis photons/pulse at the fundamental wavelength are produced. 
The flux reduction at the second (off-axis) harmonic, calculated using the theoretical model, is drastic: only about $10^8$ photons/pulse are expected at 2.5 nm. An additional factor of about $\cdot 10^4$ is lost if considering the third harmonic emission at 1.33 nm~\cite{fawley}. 
Having provided an experimental validation of the theoretical model, we can conclude that collecting off-axis emission does not allow to recover a significant photon flux when the FEL is operated in Nonlinear Harmonic Generation mode, with circularly polarized undulators.

As an alternative method to obtain significant (on-axis) circularly polarized
harmonic radiation, one may consider the possibility to operate the radiator
in elliptically polarized mode. Such a method gave interesting results in
the case of spontaneous (linear) harmonic radiation \cite{bessy}. As a
future study, we plan to investigate its efficiency when applied to
Nonlinear Harmonic Generation in FEL's.

\section{Acknowledgement}
We acknowledge useful discussions W. M. Fawley. We also thanks the Elettra machine group for the technical support.


\begin{thebibliography}{99}
\bibitem{pescia} F. Meier and D. Pescia, {\em Phys. Rev. Lett.\/} {\bf 47}, 374 (2001). 
\bibitem{blume} J. P. Hannon et al., {\em Phys. Rev. Lett.\/} {\bf 61}, 1245 (1988).   
\bibitem{altarelli} P. Carra, M. Altarelli and François de Bergevin, {\em Phys. Rev. B\/} {\bf 40}, 7324 (1989).
\bibitem{sut} J. C. Sutherland et al., {\em Nucl. Instrum. Methods Phys. Res. A\/} {\bf 172}, 195 (1994). 
\bibitem{blume2} M. Blume and D. Gibbs, {\em Phys. Rev. B \/} {\bf 37}, 1779 (1988). 
\bibitem {appl1} R. Neutze et al., {\em Nature\/} {\bf 406} 757 (2000); W. A. Barletta and H. Winck, {\em Nucl. Instr. and Meth. A\/} {\bf 500} 1 (2003); N. Gedik et al., {\em Science\/} {\bf 316} 425 (2007); A. Cavalieri, {\em Nature\/} {\bf 448} 651 (2007). 
J.T.Costello, {\em Journal of Physics: Conference Series\/} {\bf 88} 012057 (2007) and references therein, J. Kirz, {\em Nature Physics\/} {\bf 2} 799 (2006), H. N. Chapman et al., {\em Nature\/} {\bf 448} 676 (2007). 

\bibitem{Schmitt86} M.J. Schmitt and C.J. Elliott, {\em Phys. Rev. A\/} {\bf 34}, 4843 (1986).
\bibitem{Bonifacio90} R. Bonifacio, L. De Salvo and P. Pierini, {\em Nucl. Instrum. and Meth. A\/} {\bf 293}, 627 (1990).
\bibitem{pino1} G. Dattoli and G. Voykov, {\em Phys. Rev. E\/} {\bf 48}, 3030 (1993).
\bibitem{Huang00}
Z. Huang and K.J. Kim, {\em Phys. Rev. E\/} {\bf 62}, 7295-7308 (2000).
\bibitem{Biedron02}
S.G. Biedron et al., {\em Phys. Rev. ST AB\/} {\bf 5}, 030701 (2002).
\bibitem{Freund05}
H.P. Freund, G.P. O'Shea and S.G. Biedron, {\em Phys. Rev. Lett.\/} {\bf 94}, 074802 (2005).
\bibitem{Saldin06}
E.L. Saldin, E.A. Schneidmiller and M.V. Yurkov, {\em Phys. Rev. ST AB\/} {\bf 9}, 030702 (2006).
\bibitem{pino2} M. Labat et al.,  {\em  Proceeding FEL Conference 2009 \/} (http://accelconf.web.cern.ch/AccelConf/

FEL2009/papers/wepc56.pdf).
\bibitem {sase1} A. Kondratenko and E. Saldin, {\em Part. Accel.\/} {\bf 10}, 207 (1980); Y. Derbenev, A. Kondratenko, and E. Saldin, {\em Nucl. Instrum. and Meth. A\/}{\bf 193} 415 (1982); R. Bonifacio, C. Pellegrini, and L. Narducci, {\em Opt. Commun.\/} {\bf 50}, 373 (1984); M. Babzien et al., {\em Phys. Rev. E.\/} {\bf 57} 6093 (1998); S. V. Milton, {\em Phys. Rev. Lett.\/} {\bf 85} 988 (2000); J. Andruszkow, {\em Phys. Rev. Lett.\/} {\bf 85} 3825 (2000); R. Brinkmann, {\em Proceedings FEL Conference 2006\/}.
\bibitem {yu1} A. Doyuran et al., {\em Phys. Rev. Lett.\/} {\bf 86} 5902 (2001).
\bibitem {yu2} L.H. Yu et al., {\em Phys. Rev. Lett.\/} {\bf 91} 074801 (2003).
\bibitem {hhg} T. Pfeiffer et al., {\em Rep. Prog. Phys.\/} {\bf 69} 443 (2006); J. Seres at al., {\em Nature physics\/} {\bf 3} 878 (2007). 
\bibitem{prl1} E. Allaria et al., {\em Phys. Rev. Lett.\/} {\bf 100} 174801 (2008).
\bibitem{NHGP} G. Geloni, E. Saldin, E. Schneidmiller and M.
Yurkov, {\em Opt. Comm. \/} {\bf 271}, 207 (2007). 
\bibitem{OURF} G. Geloni, E. Saldin, E. Schneidmiller, M. Yurkov, {\em Opt. Comm. \/} {\bf 276}, 167 (2007). 
\bibitem{OURI} G. Geloni, E. Saldin, E. Schneidmiller, M. Yurkov, {\em Nucl. Instrum. Methods Phys. Res. A\/} {\bf 584}, 219 (2008). 
\bibitem{NHGH} G. Geloni, E. Saldin, E. Schneidmiller, M. Yurkov, {\em Nucl. Instrum. Methods Phys. Res. A\/} {\bf 581}, 856 (2007). 
\bibitem{ALFE} D. Alferov, Y. Bashmakov and E. Bessonov, {\em Sov. Phys. Tech. Phys.} {\bf 18}, 1336 (1974).
\bibitem{prl2} G. De Ninno et al., {\em Phys. Rev. Lett.\/} {\bf 101} 053902 (2008).
\bibitem{nim} C. Spezzani et al., {\em Nucl. Instrum. Methods Phys. Res. A\/} {\bf 596} 451 (2008).
\bibitem{fawley} W. Fawley, private communication.
\bibitem{fermicdr} C. Bocchetta et al., {\em FERMI@Elettra Conceptual Design Report\/}, available at ttp://www.elettra.trieste.it/FERMI.
\bibitem{bessy} S. Khan et al., {\em Phys. Rev. Lett.\/} {\bf 97} 074801 (2006).
\end {thebibliography}
\end{document}